\documentclass{article}
\usepackage[utf8]{inputenc}
\usepackage{authblk}
\usepackage[dvips]{graphicx}
\graphicspath{{noiseimages/}}
\usepackage{xcolor}
\usepackage[T2A]{fontenc}
\usepackage{tikz}
\usepackage{pgfplots}

\usepackage{mathrsfs} 
\pgfplotsset{compat=1.7}
\usetikzlibrary{intersections, pgfplots.fillbetween}
\usetikzlibrary{decorations.pathmorphing}
\usetikzlibrary{arrows.meta}
\usepackage[mode=buildnew]{standalone}
\usepackage[toc,page]{appendix}
\usepackage{amsmath}
\usepackage{amssymb}
\usepackage{float}
\usepackage{comment}
\usepackage{a4wide}
\usepackage[english]{babel}
\usepackage{hyperref}
\definecolor{refcolor}{HTML}{CD2600}
\definecolor{tablecolor}{HTML}{373641} 
\definecolor{urlcolor}{HTML}{E12900} 
\hypersetup{pdfstartview=FitH,  linkcolor=refcolor,urlcolor=urlcolor, colorlinks=true,citecolor=refcolor}
\usepackage[page]{appendix}

\author[1,2]{E.T.Akhmedov}
\author[1,2]{K.V.Bazarov\footnote{\tt bazarov.kv@phystech.edu}}
\affil[1]{Moscow Institute of Physics and Technology, Institutskii per. 9, 141700, Dolgoprudny, Russia}
\affil[2]{National Research Centre «Kurchatov Institute» }
\title{\textcolor{black}{On the backreaction issue for the black hole in de Sitter space-time}}

\begin{document}

\numberwithin{equation}{section}

\maketitle

\begin{abstract}
We consider quantum real massive scalar field in the de Sitter–Schwarzschild space-time backround. To have an analytic head way we study in detail the two-dimensional case, assuming that the situation in four dimensions will not be much different conceptually. It is assumed, that quantum field is in a thermal state i.e. described by the planckian distribution for the exact modes in the geometry under consideration. We calculate approximately the expectation value of stress-energy tensor near the cosmological and black hole horizons. It is shown that for a generic temperature backreaction from quantum fields on the geometry cannot be neglected. Thus, de Sitter–Schwarzschild space-time geometry inevitably is strongly modified by the quantum fluctuations of the matter fields.

\end{abstract}

%\tableofcontents

\section{Introduction}

There is an indirect experimental evidence that our Universe has undergone a stage of rapid inflationary expansion \cite{Planck:2018vyg,Starobinsky:1980te,Starobinsky:1982ee,Linde:1981mu,Linde:1983gd,Guth:1980zm,Guth:1982ec,Albrecht:1982wi}. It is believed that the curvature of the Universe at that stage was of the order of GUT scale. Although formation of black holes at that stage is believed to be highly unlikely, due to the rapid expansion, we think that possibility of such a formation strongly depends on the initial quantum state of the fields at the beginning of the inflation, which is not known to us so far. Furthermore, there are suggestions about hypothetical type of black holes that formed in the early Universe \cite{1966AZh....43..758Z,10.1093/mnras/152.1.75}. These so-called primordial black holes, their stability and evaporation are of interests because they are candidates for the components of the dark matter \cite{Chapline:1975ojl,1975A&A....38....5M}.

So, we would like to consider the situation when there is a black hole present during that expansion stage. This situation is modelled by the following two-dimensional part of the four-dimensional metric
\cite{1950SRToh..34..160N,Nariai1951OnAN} (see also 
\cite{Ginsparg:1982rs,Spradlin:2001pw,Cha:2002ca,Shankaranarayanan:2003ya}):
\begin{align}
\label{sdsmetric}
    ds^2=f(r) dt^2-\frac{dr^2}{f(r)}, \qquad f(r)=1-\frac{2M}{r}-H^2 r^2.
\end{align}
Here $M$ is the mass of the black hole, $H$ is Hubble constant, and $d\Omega^2$ is the line element of the unit sphere. 
%We assume that the curvature of the de Sitter space-time is parametericaly very large and, hence, quantum effects for matter fields are essential. But the gravity is still classical \textcolor{blue}{Eto on prosil procommentirovat'}.

The main feature of this metric, which is relevant for our further discussion, is the presence of two horizons simultaneously --- there is the cosmological and black hole horizon. In the present article we consider two dimensional space-time since it allows us to find the key properties of the QFT on the metric \eqref{sdsmetric} without unnecessary technical complications. However, the cost of this simplification is the absence of the dynamical gravity in two dimensions. But, we expect that the main conclusions remain true in higher dimensions, as it was in other space-times (see \cite{Akhmedov:2020ryq,Bazarov:2021rrb} to compare results in two and four dimensions).

In \cite{Kay:1988mu} it was shown that quantum field theory in such a background has certain peculiarities, which are due to the presence of two horizons. Indeed, the function $f(r)$ in (\ref{sdsmetric}) has two separated first order roots, which correspond to the presence of two horizons. But, after the Wick rotation to the Euclidian signature one encounters two different conical singularities \cite{Hawking:1976de,Gibbons:1977mu,PhysRevD.15.2752,PhysRevD.69.064016,Teitelboim:2001skl,Lin:1998pj,Lin_1999,Shankaranarayanan:2003ya}. Therefore, it is believed that this system cannot be in thermodynamic equilibrium and space-time \eqref{sdsmetric} is unstable \cite{Bousso:1996au,Bousso:1997wi,Nojiri:1998ue,Nojiri:1998ph,Medved:2002zj,PhysRevD.52.666}. See also \cite{Akhmedov:2020ryq,Akhmedov:2021agm,Akhmedov:2020qxd,Bazarov:2021rrb,Akhmedov:2021cwh} for the discussion of the physical consequences of the presence of such singularities. See also the discussion of the multi-horizon thermodynamics in e.g. \cite{Choudhury:2004ph}, where massless field was studied with an alternative approach to the problem. In the latter paper analytic continuation of the metric to the Euclidean signature was used and was climed that conical singularity can be avoided for specific relation between deficit angles of the two horizons. However, we work in the Lorentzian signature and study the backreaction due to quantum expectation value of the stress energy tensor. In general, the two approaches are not the same and the connection between them should be carefully examined in each given case.

We find it hard to judge about the destiny of quantum field theory on such a background as (\ref{sdsmetric}) without addressing the backreaction issue based on the explicit calculation of the expectation value of the stress-energy tensor (SET) and its effect on the background metric in the Lorentzian signature. The goal of the present paper is to calculate the SET. We will do the calculation for a class of states that are similar to the Hartle-Hawking state, which are believed to be stationary.

In the sections \ref{setup} and  \ref{geosec} we discuss the key ideas of the technical part of the work and take a closer look at the geometry of the space \eqref{sdsmetric} correspondingly. In the Sec. \ref{quatization} we quantize the field, and in the Sec. \ref{set2d} we find the approximate value of the SET near the horizons and discuss its properties. In the Sec. \ref{posstate} we propose a state, which may nullify the SET on both horizons. However such a state does not obey the fluctuation-dissipation theorem
\section{Setup of the problem}
\label{setup}
We consider the situation, when a quantum field is existing between two horizons, $r_c$ and $r_b < r_c$, in two dimensional space-time. Namely, in the space-time \eqref{sdsmetric} we consider the free real massive scalar field theory:
\begin{align}
\label{action}
    S=\frac{1}{2}\int d^2x \sqrt{g}\Big( \partial_\mu \varphi\partial^\mu \varphi-m^2\varphi^2\Big).
\end{align}
This two dimensional model can be considered as usual as the radial part of the four dimensional theory. To some limited extend this two dimensional theory allows to judge about the situation in the four dimensional case. In a similar situation in four dimensions the presence of the quantum scalar field leads to the appearance of the expectation value of the corresponding SET on the right hand side of the Einstein equations:
\begin{align}
\label{ein}
    G_{\mu\nu}+\Lambda g_{\mu\nu}=8\pi G \langle :\hat{T}_{\mu\nu}:\rangle,
\end{align}
where the expectation value, $\langle :\hat{T}_{\mu\nu}:\rangle$, is taken with respect to a state of the scalar field theory. Of coarse, in two dimensional space-time, there is no any dynamics of gravity, which can be described by anything like Einstein equations. However, the situation in higher dimensions is expected to be qualitatively similar. So, the two-dimensional case allows us to demonstrate the key features of the theory without unnecessary complicated calculations. The main goal of our paper is to show that independently of the choice of the state (within the class of stationary thermal states) this expectation value of the SET diverges (infinitely grows) either on cosmological or on the black hole horizon, or simultaneously on both of them. Thus, quantum field changes the background metric, which signals that there is a strong backreaction.

The metric \eqref{sdsmetric} has the time-like killing vector. Then, among the simplest possible states there is the class defined by the ``thermal'' density matrix:
\begin{align}
    \hat{\rho}=e^{-\beta \hat{H}}.
\end{align}
Where, $\hat{H}$ is the Hamiltonian of the scalar field theory under consideration. Then, the expectation value of an operator $\hat{O}$ is defined as follows:
\begin{align}
    \langle \hat{O} \rangle = \frac{\text{Tr} \hat{O} \hat{\rho}}{\text{Tr} \hat{\rho}}.
\end{align}
To calculate the expectation value of the SET we express it via the Wightman two-point function:
\begin{align}
\label{defT}
    T_{\mu\nu}(x)_\beta=\bigg(\frac{\partial}{\partial x_1^\mu}\frac{\partial}{\partial x_2^\nu}-\frac{1}{2}g_{\mu\nu}\Big[g^{\alpha\beta}\frac{\partial}{\partial x_1^\alpha}\frac{\partial}{\partial x_2^\beta}-m^2\Big]\bigg)W_\beta(x_1|x_2)\bigg|_{x_1=x_2=x}, 
\end{align}
here the Wightman function is defined as follows
\begin{align}
   W_\beta(x_1|x_2)=\langle \hat{\varphi}(x_1) \hat{\varphi}(x_2)\rangle.
\end{align}
This expectation value has the standard UV divergence which has to be regularized. 

There are different regularization methods. One of the standard methods is to use the covariant point-splitting
\cite{PhysRevD.13.2720,Bunch:1978yq,1977RSPSA.354...59D,PhysRevD.51.4337} (see also \cite{Diatlyk:2020nxa,Akhmedov:2020ryq,Akhmedov:2021agm}). In fact, the key feature of the two-dimensional case is the following: the behavior of the modes at horizon can be expressed in terms of in-going and out-going plane waves, which are multiplied by the transition and the reflection coefficients calculated in the corresponding effective potential due to the background metric. We show this explicitly below. 

Moreover, for the class of the states of the field under consideration \textcolor{black}{the leading contribution to the SET near the horizon do not depend on the transition and the reflection coefficients separately}. Terms which depend on these coefficients contribute in the special combination, as we will see in \eqref{t002d}. It turns out, that this combination is nothing but the sum of the probability of the transition and reflection, which always is equal to one. Hence, in two dimensions one can calculate the behavior of the SET on the horizons without knowing expressions of the reflection and transition coefficients. As a result, in two dimensions for the states that we consider the  expectation value contains two terms. The first term depends only on the temperature and does not depend on the geometry. While the second term, which comes from regularization does depend only on geometry, but does not depend on the state of field.

Also we want to stress here that in such a background as we consider here the expectation value depends on the spatial coordinate $r$ and cannot be found exactly for any $r$. However, we can calculate the approximate expectation value of the SET near the horizons, where its behaviour is the most interesting for the backreaction problem. 
\section{The geometry}
\label{geosec}
In this section we discuss the geometry of the space-time with the metric \eqref{sdsmetric}. The function $f(r)$ has two zeros for $r\geq 0$: one, $r_c$, is corresponding to the cosmological horizon, while the other, $r_b \leq r_c$, --- to the black hole horizon. In (\ref{sdsmetric}) the range of $r$ is as follows: $r_b\le r\le r_c$. In the situation under consideration $f(r)$ can be written as:
\begin{align} \label{f}
    f(r)=H^2\frac{(r-r_b)(r_c-r)(r+r_c+r_b)}{r}.
\end{align}
If we were considering the 4D case the positions of the horizons, $r_c$ and $r_b$, could be related to the Hubble constant and black hole mass as follows: 
\begin{align*}
    H^2=\frac{1}{r_c^2+r_b r_c +r_b^2},\qquad M=\frac{r_b r_c(r_b+r_c)}{2(r_c^2+r_b r_c+r_b^2)}=H^2\frac{r_b r_c(r_b+r_c)}{2}.
\end{align*}
The condition $r_b \leq r_c$ requires the following restriction on the black hole mass:
\begin{align*}
    M \leq \frac{1}{3\sqrt{3}H},
\end{align*}
where the equality corresponds to the extremal case $r_b=r_c$.

Note that the case $r_c\to\infty$ corresponds to the asymptotically flat Schwarzschild solution. At the same time $r_b=0$ corresponds to the empty de Sitter space-time $r_c=H^{-1}$. However, one needs to be careful in taking such limits. In fact, for example, the space-time with non-zero $M$ (but even very small) is topologically different from the space with $M=0$.

To quantize a field theory in such a space-time, it is necessary to impose boundary conditions on the horizons. For these purposes it is more convenient to use conformal variant of the metric \eqref{sdsmetric}. Namely, one can define \cite{PhysRev.110.965}:
\begin{align}
\label{conformal}
    ds^2=f(r^*)\big(dt^2-dr^*{}^2\big), \quad {\rm where} \quad r^*=\int \frac{dr}{f(r)}.
\end{align}
As usual to express $r$ via $r^*$ one has to solve the following transcendental equation:
\begin{align}
\label{rsr2d}
    r^*=-\frac{r_b \log (r-r_b)}{H^2 (r_b-r_c) (2 r_b+r_c)}+\frac{r_c \log (r_c-r)}{H^2 (r_b-r_c) (r_b+2 r_c)}+\frac{(r_b+r_c) \log (r+r_b+r_c)}{H^2 (2 r_b+r_c) (r_b+2
   r_c)}.
\end{align}
It is curious that for a specific values of $r_{b,c}$:
\begin{align}
\label{specset}
    r_b=\frac{\sqrt{3}-1}{2}, \quad r_c=1,
\end{align}
one can solve the transcendental eq. \eqref{rsr2d} explicitly
\begin{align}
    r=r_b \frac{\left(1+\sqrt{3}\right) \left(2 e^{2r^*}+1\right)-\left(3+\sqrt{3}\right) \sqrt{2 e^{2r^*}+1}}{2 \left(e^{2r^*}-1\right)}.
\end{align}
For generic values of $r_{b,c}$ the equation can be solved only approximately in the near horizon limits:
\begin{align*}
   r\approx r_b + e^{\frac{H^2 r^* (r_c-r_b) (2 r_b+r_c)}{r_b}}\quad\text{as}  \quad r \to r_b, \, r^*\to-\infty,
\end{align*}
and
\begin{align*}
  r\approx r_c-e^{-\frac{H^2 r^* (r_c-r_b) (r_b+2 r_c)}{r_c}}\quad \text{as} \quad r \to r_c, \, r^*\to+\infty.
\end{align*}
Correspondingly:
\begin{align}
\label{approxf}
    f(r*)\approx \frac{H^2(2r_b+r_c)(r_c-r_b)}{r_b}e^{\frac{H^2 r^* (r_c-r_b) (2 r_b+r_c)}{r_b}}\quad\text{as} \ \ r^*\to-\infty, 
\end{align}
and
\begin{align}
\label{approxf2}
f(r^*)\approx\frac{H^2(r_b+2r_c)(r_c-r_b)}{r_c}e^{-\frac{H^2 r^* (r_c-r_b) (r_b+2 r_c)}{r_c}} \quad\text{as} \ \ r^*\to+\infty.
\end{align}
Thus, one can define the canonical (inverse) temperatures due to the black hole and cosmological horizons:
\begin{align}
\label{betas}
    \beta_b=\frac{4\pi r_b}{H^2(r_c-r_b)(2r_b+ r_c)} = \frac{2\pi}{\kappa_b},\quad \beta_c=\frac{4\pi r_c}{H^2(r_c-r_b)(r_b+ 2r_c)} = \frac{2\pi}{\kappa_c}.
\end{align}
Here $\kappa_b$ and $\kappa_c$ are the surface gravities on the event and cosmological horizons, correspondingly. One of the important properties of the geometry under consideration is that these temperatures are not equal for any values of $r_b,r_c$:
\begin{align}
    \beta_c-\beta_b=\frac{2\pi(r_b+r_c)}{H^2(2r_b+r_c)(r_c+2r_b)}>0
\end{align}
Furthermore, note that if $r_b\to r_c$, then both $\beta_{b,c} \to \infty$, but their difference remains finite.  %This fact follows from the following: if $r_b\ne r_c$, then the both radii are first order roots of the $f(r)$. But, if $r_c=r_b$, merge together and second order root of the $f(r)$ appears.

Below we will show, that in the two dimensional case, to calculate the SET in the vicinity of the horizons we need to know only the behavior of $f(r)$ and of the modes near the horizons.

\section{Quantization}
\label{quatization}
The equations of motion for the scalar field with the action eq. \eqref{action} in such a background gravitational field as \eqref{conformal} are as follows:
\begin{equation}
   \Big[ -\partial_{r^*}^2+m^2f(r^*)\Big]\varphi_\omega(r^*)=\omega^2\varphi_\omega(r^*), \quad {\rm where} \quad \varphi(t,r^*)=\frac{1}{\sqrt{2\omega}}e^{i\omega t}\varphi_\omega(r^*). \label{twoRs}
\end{equation}
Thus, in the radial direction one obtains a quantum mechanical scattering problem. As a result, the full basis of solutions consists of out-going $\overrightarrow{\varphi}_\omega(r^*)$ and in-going $\overleftarrow{\varphi}_\omega(r^*)$ modes with the following behaviour near horizons\footnote{Note that the coefficients $R_\omega$ depend on the scattering direction and are related to each other as follows: $\overrightarrow{R}_\omega=-\overleftarrow{R}_\omega^*\frac{T_\omega}{T^*_\omega}$, while $T_\omega\equiv\overrightarrow{T}_\omega=\overleftarrow{T}_\omega$. However below we will need only absolute values of these coefficients, which obey the following condition: $|R_\omega|\equiv|\overrightarrow{R}_\omega|=|\overleftarrow{R}_\omega|$. %Furthermore, to calculate the stress-energy tensor expectation value we will need zero energy limit of these coefficients. But, as we will see below $\overrightarrow{R}_0=\overleftarrow{R}_0$. 
That is the reason why we do not distinguish between the coefficients $\overrightarrow{R}_\omega $and $\overleftarrow{R}_\omega$ in eq. (\ref{twoRs}) and below.}:
\begin{center}
\begin{tabular}{|c|c|c|}
\hline
\ &  $r^*\to-\infty $& $r^*\to+\infty$ \\  \hline
$\overrightarrow{\varphi}_\omega(r^*)$  & $e^{i\omega r^*}+R_\omega e^{-i\omega r^*}$ & $ T_\omega e^{i\omega r^*} $ \\ \hline
$\overleftarrow{\varphi}_\omega(r^*) $ & $T_\omega e^{-i\omega r^*} $&  $e^{-i\omega r^*}+R_\omega e^{i\omega r^*}$\\
\hline
\end{tabular}
\end{center}
\ \\
There are the following obvious condition for the reflection and transition coefficients: 
\begin{align}
    \label{RTcond}
    |T_\omega|^2+|R_\omega|^2=1,
\end{align}
which follows from the normalization conditions that can be related to the canonical commutation relations between the field operator and its conjugate momentum and between creation and annihilation operators. The field operator has the following form:
\begin{align}
\label{fieldopertaro}
    \hat{\varphi}(t,r^*)=\int_0^\infty\frac{d\omega}{\sqrt{2\pi}}\frac{1}{\sqrt{2\omega}}\bigg[e^{i\omega t}\Big(\hat{a}^\dagger_\omega \overrightarrow{\varphi^{}}_\omega(r^*) +\hat{b}^\dagger_\omega \overleftarrow{\varphi^{}}_\omega(r^*) \Big)+h.c. \bigg],
\end{align}
where the creation and annihilation operators obey the standard commutation relations:
\begin{align}
    [\hat{b}^{}_\omega,\hat{b}^\dagger_{\omega'}]=\delta(\omega-\omega'),\qquad [\hat{a}^{}_\omega,\hat{a}^\dagger_{\omega'}]=\delta(\omega-\omega'), \qquad [\hat{a}^{}_\omega,\hat{b}^\dagger_{\omega'}]=0.
\end{align}
Then the thermal state corresponds to the following expectation values:
\begin{align}
    \hat{\rho}=e^{-\beta \hat{H}}, \ \ \text{then} \ \ \langle \hat{a}^\dagger_\omega,\hat{a}^{}_{\omega'}\rangle=\frac{1}{e^{\beta \omega}-1}\delta(\omega-\omega'), \ \ \langle \hat{b}^\dagger_\omega,\hat{b}^{}_{\omega'}\rangle=\frac{1}{e^{\beta \omega}-1}\delta(\omega-\omega').
\end{align}
So one obtains the Wightman function as follows:
\begin{align}
\label{wbeta}
    W(t_2,r_2^*|t_1,r_1^*)=\int_{-\infty}^{\infty} \frac{d\omega}{2\pi} \frac{e^{i\omega(t_2-t_1)}}{e^{\beta\omega}-1}\frac{1}{2\omega}\Big[ \overrightarrow{\varphi^{}}_\omega(r_1^*) \overrightarrow{\varphi^*}_\omega(r_2^*)+\overleftarrow{\varphi^{}}_\omega(r_1^*) \overleftarrow{\varphi^*}_\omega(r_2^*)\Big],
\end{align}
which we will use to calculate the SET expectation value. Please note the integration limits in \eqref{wbeta} and in \eqref{fieldopertaro}. The same quantization procedure was discussed, for example, in \cite{Anderson:2022icz,Anderson:2022zuc,Firouzjahi:2022vij,Singha:2021pkg}. 
\section{Stress-energy tensor}
\label{set2d}
Let us start with the calculation of the energy density near say the black hole horizon (see also \cite{10.1143/PTP.83.941}). The relevant behaviour of the Wightman function near the corresponding horizon is:
\begin{multline}
\label{appW2d}
    W(t_2,r_2^*|t_1,r_1^*)\approx\int_{-\infty}^{\infty} \frac{d\omega}{2\pi} \frac{e^{i\omega(t_2-t_1)}}{e^{\beta\omega}-1}\frac{1}{2\omega}\Big[e^{i\omega(r_2^*-r_1^*)}+|R_\omega|^2e^{-i\omega(r_2^*-r_1^*)}+\\+R_\omega e^{i\omega(r_1^*+r_2^*)}+R^*_\omega e^{-i\omega(r_1^*+r_2^*)}+ |T_\omega|^2e^{-i\omega(r_2^*-r_1^*)}\Big].
\end{multline}
The approximate form of the Wightman function near the cosmological horizon is very similar. Note, that near the horizon the Wightman function contains two types of terms under the integral on the RHS of \eqref{appW2d}. The first type depends on $r_2^*-r_1^*$ while the second type depends on $r_1^*+r_2^*$. By definition \eqref{defT} the energy density can be expressed via the Wightman function as:
\begin{align}
\label{t002d}
    \langle T_{00} \rangle=\frac{1}{2}\Big[\frac{\partial}{\partial_{t_1}}\frac{\partial}{\partial_{t_2}}+\frac{\partial}{\partial_{r_1^*}}\frac{\partial}{\partial_{r_2^*}}\Big]W(t_2,r_2^*|t_1,r_1^*)\Big|_{1\to2}.
\end{align}
Due to the structure of derivatives in the eq. \eqref{t002d} the terms, which depend on $r_1^*+r_2^*$ in \eqref{appW2d} do not contribute to the $T_{00}$ near the horizon, and we obtain that:
\begin{align}
\label{t002d2}
    \langle T_{00}\rangle\approx \int_{-\infty}^{\infty} \frac{\omega d\omega}{4\pi} \frac{1}{e^{\beta\omega}-1}\Big[1+|R_\omega|^2+|T_\omega|^2\Big]=\int_{-\infty}^{\infty} \frac{\omega d\omega}{2\pi} \frac{1}{e^{\beta\omega}-1}.
\end{align}
Where we have used the properties of the reflection and transition coefficients \eqref{RTcond}. In \eqref{t002d2} there is the standard UV divergence due to the zero-point fluctuations. As we agreed above we use the covariant point splitting method to obtain that (see the similar calculations in \cite{Akhmedov:2020ryq,Diatlyk:2020nxa,Akhmedov:2021agm}): 
\begin{align}
\label{Tresult}
    \langle :T_{00}:\rangle\approx \frac{\pi}{6}\frac{1}{\beta^2}-\frac{1}{6\pi} f(r^*)^{1/2}\frac{\partial^2}{\partial^2r^*}f(r^*)^{-1/2}.
\end{align}
Here we presented only $T_{00}$ component and omitted details of regularization. For more details and all other components of SET see App.\ref{appA}. Concerning $f(r^*)$, note that it has different approximate behaviour near the two horizons: \eqref{approxf} and \eqref{approxf2} correspondingly. Hence, the energy density near the black hole horizon $r^*\to-\infty$ and near the cosmological horizon $r^*\to\infty$ has different asymptotics, respectively:
\begin{align}
    \begin{matrix}
\text{Black hole horizon} \qquad& \langle :T_{00}:\rangle\approx\frac{\pi}{6}\Big[\frac{1}{\beta^2}-\frac{1}{\beta_b^2}\Big] \\
\text{Cosmological horizon} \qquad & \langle :T_{00}:\rangle\approx\frac{\pi}{6}\Big[\frac{1}{\beta^2}-\frac{1}{\beta_c^2}\Big].
\end{matrix}
\end{align}
Absolutely similarly one can obtain $\langle :T_{11}:\rangle$ near both horizons. Recall that $\beta_c\ne\beta_b$ according to \eqref{betas}. For this reason, we cannot choose any thermal state that nullifies the SET near the both horizons simultaneously. Finite expectation value of the SET at least on one of the horizons signals the presence of strong backreaction. Since the metric tensor degenerates near the horizon it is very sensitive even to small perturbations. \textcolor{black}{For the more details see appendix \ref{appB}.}

On the one hand, this reasoning doesn't make much sense in two dimensions, because the gravitational field is not dynamical. On the other hand, in two dimensions, we obtained the SET by a much simpler calculations than in four dimensions. In the last section we conclude with the physical predictions for the backreaction problem in four dimensions.

\section{Possible candidates for the state with the zero SET expectation value near the horizons}
\label{posstate}
In the previous section, we have considered only thermal states. It was shown, that for any $\beta$ the expectation value of the SET is not zero at least on one of the two horizons. However, if we go beyond the class of thermal states, we can find the state, which corresponds to the zero SET near both horizons. In fact, consider the following state:
\begin{align}
    \langle \hat{a}^\dagger_\omega,\hat{a}^{}_{\omega'}\rangle=n_{out}(\omega)\delta(\omega-\omega'), \qquad \langle \hat{b}^\dagger_\omega,\hat{b}^{}_{\omega'}\rangle=n_{in}(\omega)\delta(\omega-\omega'), \qquad \langle \hat{b}^\dagger_\omega,\hat{a}^{}_{\omega'}\rangle=0.
\end{align}
Where $n_{out}(\omega)$ and $n_{in}(\omega)$ are the distributions of the out-going and in-going modes, which are generic functions of $\omega$, not necessarily equal to the planckian distribution.
Let us take them in the following form:
\begin{align}
    n_{out}(\omega)=\frac{1}{e^{\beta_0\omega}-1}+\delta n(\omega), \quad n_{in}(\omega)=\frac{1}{e^{\beta_0\omega}-1}-\delta n(\omega), \quad\text{where}\quad  \frac{2}{\beta_0^2}=\frac{1}{\beta_b^2}+\frac{1}{\beta_c^2},
\end{align} 
where $\delta n$ satisfies the following condition $\delta n(-\omega)=-\delta n(\omega)$. Note also that $\beta_0$ does not coincide with the effective temperature of the Schwarzshild de Sitter space \cite{Shankaranarayanan:2003ya}. We propose to consider a class of states which are close to the thermal one, and $\delta n$ plays the role of the difference in the number of the out-going and the in-going particles, $n_{out}(\omega)-n_{in}(\omega)=2\delta n(\omega)$.

Because, $n_{out}(\omega)\ne n_{in}(\omega)$ in general there should be present a non-zero energy flux $T_{01}$. To avoid such apparent non-stationary situation, we obtain two restricting equations on $\delta n(\omega)$. The first one comes from the conditions that $\langle : T_{00}: \rangle=0 $ and $\langle : T_{11}: \rangle=0$, while the second one -- comes from the condition $\langle : T_{01}: \rangle=0$. Using expressions similar to \eqref{t002d2}, but with $n_{out}(\omega) \neq n_{in}(\omega)$, one can find the following restrictions on $\delta n(\omega)$:
\begin{align}
\label{dneq}
    \int_0^\infty d\omega \omega \delta n(\omega) |R_\omega|^2=\frac{\pi^2}{12}\Big[\frac{1}{\beta_b^2}-\frac{1}{\beta_c^2}\Big], \qquad \int_0^\infty d\omega \omega \delta n(\omega) |T_\omega|^2=0.
\end{align}
The question is whether there is such a solution of these equations which corresponds to a stationary state --- which obeys the fluctuation-dissipation theorem. In two dimensions, due to the specifics of the scattering processes there, it is quite plausible to find such a state. But in 4D we do not think that any of such states is stationary.
\section{Conclusions}
It is stated in many papers \cite{Bousso:1996au,Bousso:1997wi,Medved:2002zj,PhysRevD.52.666} that it is impossible to achieve the thermal equilibrium in Schwarzshild de Sitter space-time. To understand the latter statement deeper, we study quantum field theory on such a background. Our conclusions are opposite -- we think there is n equilibrium state, but it strongly affects the backround.

We show that there is no thermal state for which the expectation value of the SET vanishes on both horizons. On the contrary, the expectation value either blows up on one of the horizons or on both of them simultaneously, depending on the choice of the temperature. 

Let us present here a few more comments. Despite the fact, that $T_{\mu\nu}$ is finite on the horizon in the coordinate system under consideration, the $T_\mu^\nu$ tensor blows up there. Essentially in this article we consider the two dimensional situation as the radial part of the four dimensional one (see e.g. for a similar discussion \cite{Bazarov:2021rrb}). Hence, in four dimensional case with such a SET one would encounter the situation that left hand side of the Einstein equations \eqref{ein} will be finite, while the right hand side will diverge. This signals the strong backreaction on the gravitational background under consideration. The situation is similar to the quantum field theory in black hole background in the Boulware state. We adopt here the point of view that quantum field theory can exist in any background in any Hadamard state. Just in the Boulware state the backrecation is so strong that it eliminates the black hole horizon \cite{Fabbri:2005zn,Ho:2018fwq,Ho:2018jkm}. Meanwhile in the black hole de Sitter background for any state, at least within the class of thermal ones, we see that the backreaction is strong on either black hole or cosmological horizon. E.g., if we consider the state with the temperature of the cosmological horizon, then expectation value of the SET on the black hole horizon blows up and eliminates it. I.e. for such a state one will obtain a geometry without BH horizon, which is similar to \cite{Fabbri:2005zn,Ho:2018fwq,Ho:2018jkm}. 

Furthermore, we expect that a collapse of a matter into a the black hole in the de Sitter space may follow a completely different scenario than in the case of the absence of the cosmological horizon, if one takes into account quantum fluctuations. 

\section*{Acknowledgments}
We would like to acknowledge valuable discussions with D.V.Diakonov and P.A.Anempodistov. This work was supported by the grant from the Foundation for the Advancement of Theoretical Physics and Mathematics ``BASIS'' and by the Russian Ministry of education and science.
\begin{appendices} 
\numberwithin{equation}{section}
\setcounter{equation}{0}
\section{Regularization \label{appA}}
In the calculation of the expectation value of the SET one has to regularize it. In our work we used point splitting method \cite{Christensen:1978yd,Bunch:1978yq}. More recently similar calculations were performed e.g. in \cite{Akhmedov:2020ryq,Diatlyk:2020nxa,Akhmedov:2021agm}. To make the paper self-contained and to set up the notations, in this appendix we summarize the standard point splitting regularization procedure of the expectation value of the stress--energy tensor in curved space--time with the metric of the form:
\begin{align}
    ds^2=C(u,v)dudv
\end{align}
The SET is given by the following expression:
\begin{align*}
    \langle \hat{T}_{\mu\nu}(x) \rangle = \left. D_{\mu\nu} \langle \hat{\varphi}(x^+)\hat{\varphi}(x^-) \rangle\right|_{x^+=x^-=x}.
\end{align*}
Here $D_{\mu\nu}$ is a differential operator; $x^\pm$ are points which are separated from $x$ along a geodesic with tangent vector $t^\mu$. A point close enough to $x^\mu$ can be represented as follows  
\begin{align}
\label{eqgeo}
    x^\mu(\tau)=x^\mu+\tau t^\mu+\frac{1}{2} \tau^2 a^\mu+\frac{1}{6}\tau^3 b^\mu+..., 
\end{align}
where $\tau$ is the proper length, and $a^\mu$ and $b^\mu$ can be found from the geodesic equation. Straightforward calculation gives for the thermal state the following result:
\begin{align}
    \langle   T_{\mu \nu}\rangle = - \bigg[ \frac{1}{4 \pi \epsilon^2 (t_{\alpha} t^{\alpha})} + \frac{R}{24 \pi} \bigg] \bigg[ \frac{t_{\mu} t_{\nu}}{t_{\alpha} t^{\alpha}} - \frac{1}{2} g_{\mu \nu} \bigg] + \Theta_{\mu \nu}.
\end{align}
Thus the regularized SET reads:
\begin{align}
\label{regTapp}
    \langle   :T_{\mu \nu}: \rangle =\Theta_{\mu \nu} + \frac{R }{48 \pi } g_{\mu \nu},
\end{align}
with
\begin{align}
    \Theta_{uu} &= -\frac{1}{12 \pi} C^{1/2} \partial^2_u C^{-1/2}+ \text{state dependent terms},\\
    \Theta_{vv} &= -\frac{1}{12 \pi} C^{1/2} \partial^2_v C^{-1/2} + \text{state dependent terms},\\
    \Theta_{uv} &= \Theta_{vu} =0.
\end{align}
The second term in \eqref{regTapp} is the same as in the conformal anomaly. We do not include this term in our results because it can be absorbed into the renormalization of the cosmological constant. We are interested in the first term in \eqref{regTapp}. It has two contributions: the first comes from the geometry and is independent of the state and mass of the field (and therefore does not depend on the reflection and transmission coefficients). The second one depends on the state and, in general, can depend on the reflection and transmission coefficients. But as we show by the explicit calculation these coefficients are included into the SET within a special combination in such a way that the final result for SET does not depend on them separately.

\section{Free falling reference system \label{appB}}

In this appendix we show that the SET which is found in the main body of the paper leads to the singularity in the free falling reference frame. 

To begin with, let us define the free falling reference system, which is similar to the Lemaître coordinates in the Schwarzschild black hole space-time. Let us define the time-like $\tau$ and space-like $\rho$ coordinates as follows:
\begin{align}
    d\tau=dt+[1-f(r^*)]^{\frac12}dr^*, \quad d\tau=dt+[1-f(r^*)]^{-\frac12}dr^*.
\end{align}
In these coordinates the 2D Schwarzschild de Sitter metric takes the following form:
\begin{align}
    ds^2=f(r^*)(dt^2-dr^{*2})=d\tau^2-[1-f(r^*)]d\rho^2.
\end{align}
The coordinate transformation from the coordinates $x^\mu=(t,r^*)$ to the coordinates $\tilde{x}^\mu=(\tau,\rho)$ can be described by the following matrix:
\begin{align}
    \frac{\partial \tilde{x}^{\tilde{\mu}}}{\partial x^\nu}=\Lambda^{\tilde{\mu}}_{\ \nu}=\begin{pmatrix} 1 & [1-f(r^*)]^{\frac12} \\ 
    1 & [1-f(r^*)]^{-\frac12}\end{pmatrix}.
\end{align}
From the eq. \eqref{Tresult}, the stress-energy tensor near the horizons in $(t,r^*)$ coordinates is given by:
\begin{align}
\label{appbeq4}
    T^{\mu\nu} \approx A \frac{1}{f(r^*)^2} \begin{pmatrix}
        1 & 0 \\ 0 & 1
    \end{pmatrix},
\end{align}
where:
\begin{align}
    \begin{matrix}
\text{Near black hole horizon} \qquad& A=\frac{\pi}{6}\Big[\frac{1}{\beta^2}-\frac{1}{\beta_b^2}\Big] \\
\text{Near cosmological horizon} \qquad & A=\frac{\pi}{6}\Big[\frac{1}{\beta^2}-\frac{1}{\beta_c^2}\Big].
\end{matrix}
\end{align}
The most important point here is that $A$ is not zero at least on one of the horizons. It cannot be made zero on both of them simultaneously. Hence, \eqref{appbeq4} has the following form in the $(\tau,\rho)$ coordinates:
\begin{align}
    T_{\tilde{\mu}\tilde{\nu}}=g_{\tilde{\mu}\tilde{i}}g_{\tilde{\nu}\tilde{j}}T^{\tilde{i}\tilde{j}}=g_{\tilde{\mu}\tilde{i}}g_{\tilde{\nu}\tilde{j}}\Lambda^{\tilde{i}}_{\ i}\Lambda^{\tilde{j}}_{\ j}T^{ij}=\frac{A}{f(r^*)^2} \begin{pmatrix}
        2-f(r^*) & 2(f(r^*)-1) \\ 2(f(r^*)-1)& (2-f(r^*))(1-f(r^*))
    \end{pmatrix}.
\end{align}
Near the horizons $f(r^*)$ tends to zero, so $T_{\tilde{\mu}\tilde{\nu}}$ is singular. The four dimensional counterpart of this situation would lead to the case that the geometrical part of Einstein equations is constant while the stress-energy tensor part is divergent. Now we see this is true at least for the radial part of the problem, i.e. if the angles can be ignored.

\end{appendices}
\renewcommand\theequation{A.\arabic{equation}}
\bibliographystyle{unsrturl}
\bibliography{bibliography.bib}
\end{document}